\documentstyle[sprocl]{article}

\def\Journal#1#2#3#4{{#1} {\bf #2}, #3 (#4)}

\def\NPB{{\em Nucl. Phys.} B}
\def\PLB{{\em Phys. Lett.}  B}
\def\PRL{\em Phys. Rev. Lett.}
\def\PRD{{\em Phys. Rev.} D}

\def\be{\begin{equation}}
\def\ee{\end{equation}}
\def\bea{\begin{eqnarray}}
\def\eea{\end{eqnarray}}

\begin{document}

\title{HIGH TEMPERATURE SYMMETRY NONRESTORATION}

\author{BORUT BAJC}

\address{Department of Physics, New York University, 
New York, NY 10003, USA\\E-mail: bb46@is2.nyu.edu
\\and\\
J. Stefan Institute, Jamova 39, 1001 Ljubljana, Slovenia
\\E-mail: borut.bajc@ijs.si} 


\maketitle\abstracts{ 
This is a short review on the subject of symmetry nonrestoration 
at high temperature. Special emphasis is put on experimental 
discoveries and different theoretical mechanisms. At the end, 
possible cosmological applications are briefly mentioned.}

\section{Introduction}\label{intro}

Naively one expects that at low temperature a system has less 
symmetry than at high temperature. However, there are cases in 
nature, where the opposite happens. This phenomenon is called 
inverse symmetry breaking. 

A similar phenomenon is 
symmetry nonrestoration at high temperature. It appears 
when the system has at high $T$ less symmetry than allowed 
by the Lagrangian (some vev is nonzero). 
With high temperature we mean temperature 
higher than any parameter of mass dimension in the Lagrangian. 

In this short review we will describe some examples of this 
phenomenon both in nature and in field theory. 
Due to lack of space many topics will be 
mentioned only briefly. The interested reader can refer to 
some other reviews \cite{reviews} as well as 
to the original papers. 

\section{Experimental signatures}\label{exp}

We will describe here two examples in nature, which exhibit 
the strange phenomenona of symmetry nonrestoration or 
inverse symmetry breaking. 

{\bf The Rochelle Salt.} The system \cite{jona62,weinberg74} 
has in the interesting regime two critical 
temperatures. Below $T_{c1}=-18^o$ C and above $T_{c2}=24^o$ C 
the unit cell of the Rochelle salt is orthorhombic, while it is 
monoclinic in between. Since the orthorhombic unit cell is 
more symmetric than the monoclinic one, it is the phase 
transition at the lower critical temperature $T_{c1}$ to be 
counterintuitive: heating the system we get a less symmetric 
object. Of course the next phase transition at $T_{c2}$ 
restores again the symmetry, and symmetry nonrestoration is 
thus present only in the interval between $T_{c1}$ and $T_{c2}$. 

{\bf Liquid Crystals (SmC$^*$).} The second example of 
inverse symmetry breaking, or, as it is 
called in condensed matter, of re-entrant phase behavior, is 
relatively recent. The existence of the phase was argued in 1995 
\cite{cepic95} and later experimentally found in 1998 \cite{mach98}. 

First, why is the system called SmC$^*$? Sm stays for smectic, 
i.e. with layers. The system is made from elongated molecules 
grouped in layers. 
C stays for tilted, which means that the long axis of the 
molecules form a nonzero angle with respect to 
the normal of the layers. Finally, the star $^*$ means that the 
molecules are chiral.

Second, why is SmC$^*$ a liquid crystal? The molecules in the same 
layer are behaving like a liquid, since there is no positional 
order and they can freely move in the plane. 
There is however an orientational order, since all the molecules 
in the same layer point toward the same direction. 

Clearly, there are two angles which describe the direction of 
each molecule, the tilt angle $\theta$ (the angle between the 
direction of the molecule and the normal to its layer) and the
azimuthal angle $\phi$ (the angle between the projection of 
the molecule's direction on the layer's plane and a specified 
fixed direction in the same plane). As we said before, the 
tilt angle $\theta$ is fixed for all the molecules in the 
whole liquid crystal. As regarding the azimuthal angle $\phi$, 
it is equal for all the molecules in the same layer, but it 
differs from layer to layer. However, the 
difference of this angle between any two neighbor layers is 
fixed in the whole system, i.e. the difference $\alpha=
\phi_{j+1}-\phi_j$ for layers $j+1$ and $j$ does not depend 
on the choice of $j$. So, the whole SmC$^*$ liquid crystal 
can be described by two angles, $\theta$ and $\alpha$, 
which are the order parameters of the system. 

The system is in a crystal phase below $T\approx 110^o$ C, 
while above $T\approx 120^o$ C the tilt angle $\theta\approx 0$, 
so that $\alpha$ is not defined. The interesting regime is thus 
between these two temperatures, where the polar angle is 
constant, $\theta\approx 20^o$. 

What is important is the behavior of the azimuthal 
angle as function of the temperature, $\alpha(T)$. It comes out 
that $\alpha(T)$ varies in this range, and that at different 
temperatures it changes abruptly and discontinuously. This 
signals first order phase transitions. So the system undergoes 
through different phases. The interesting point is that this 
change is not monotone: rising the temperature one goes 
through phases with $\alpha\ne 0$ to a phase with $\alpha=0$ and 
later to a phase with $\alpha\ne 0$ again. This is clearly 
similar to the case of the Rochelle salt and the same 
conclusions can be applied also here.

\section{Field Theory}\label{theory}

The known examples of symmetry nonrestoration in field 
theory can be divided into three different classes, 
which will be described below.

{\bf The prototype case.} This case was first studied by Weinberg
\cite{weinberg74} 
and later on by Mohapatra and Senjanovi\' c \cite{mohapatra79}, 
who were the first to recognize the important phenomenological 
applications of the phenomenon of symmetry nonrestoration at 
high temperature. 

The simplest model consists of two real scalar fields and a 
$Z_2\times Z_2$ symmetry, with the zero temperature potential 
given by

\begin{equation}
V={\lambda_1\over 4}\phi_1^4+{\lambda_2\over 4}\phi_2^4
+{\lambda\over 2}\phi_1^2\phi_2^2+{\mu_1^2\over 2}\phi_1^2+
{\mu_2^2\over 2}\phi_2^2\;.
\end{equation}

The boundedness from below of this potential requires 
that $\lambda_{1,2}>0$ and $\lambda_1\lambda_2>\lambda^2$. 
At high temperature ($T>>\mu_{1,2}$) one uses the general 
one-loop formula to calculate the leading correction:

\begin{equation}
\Delta V_T={T^2\over 24}\sum_i{\partial^2 V\over\partial\phi_i^2}=
{T^2\over 24}(3\lambda_1+\lambda)\phi_1^2+
{T^2\over 24}(3\lambda_2+\lambda)\phi_2^2\;.
\end{equation}

One can now choose the parameters so that 
$(3\lambda_1+\lambda)<0$ and obtain a nonzero vev for 
the first field, $<\phi_1>\ne 0$, spontaneously breaking 
in this way the first discrete symmetry $Z_2$. 
Due to the boundedness conditions the same can not be done 
for the second $Z_2$. In fact $(3\lambda_2+\lambda)$ must now 
be positive, so the second field does not develop a nonzero vev, 
i.e. $<\phi_2>=0$. The reason for the idea to work is the 
choice of a large negative Higgs coupling ($\lambda$).

Since at a very high temperature, the temperature itself is 
the only mass scale in the problem, the nonzero vev of $\phi_1$ 
must be proportional to the temperature. So, we have an example of 
symmetry nonrestoration, which persists at arbitrary high temperatures. 

What happens with higher order terms or nonperturbative 
contributions? 
In the case of global symmetries different techniques have shown 
that symmetry nonrestoration is a possible phenomenon 
(see however some opposite claims \cite{globalno}), but that 
the parameter space where this can happen tends to be smaller 
than at one loop \cite{globalyes,lattice}. 
In the case of gauge symmetries many calculations indicates 
the opposite, i.e. that for physical values of the gauge 
couplings symmetry restoration is probably unavoidable 
\cite{mohapatra79,dms95,localno}.

{\bf Flat directions.} This case most 
naturally happens in supersymmetry, which is 
particularly welcome, since the trick of the previous section 
cannot be applied to susy models. In fact 
there is a no-go theorem \cite{habermangano}, 
which states that at high enough temperature 
any internal symmetry gets always restored in renormalizable 
susy models. This is because the coupling 
constants are much more constrained in 
susy than in ordinary models, so that the small 
island of parameter space which allows symmetry nonrestoration 
in ordinary theories completely disappears when one looks at 
the supersymmetric subspace. The same seems to be 
true also in nonrenormalizable susy models 
\cite{bmsdt}. 

To avoid the no-go theorem one can consider a field which 
is not in thermal equilibrium with the rest of the system
\cite{flat98}, 
i.e. its interaction is negligible compared to the expansion 
rate of the universe. This means essentially that its coupling 
must be suppressed by inverse powers of a large cutoff. 
For this reason it does not get necessarily a positive high temperature 
mass term, which is the main reason for symmetry restoration. 

In ordinary nonsupersymmetric 
models nothing forbids a term $|\phi|^4$, 
which would again put the field $\phi$ in equilibrium with 
the system. Here is where supersymmetry plays its role. It 
can not only easily forbid dangerously strong terms, but it 
is even very natural to have plenty of flat or quasi-flat 
directions, which are not coupled or extremely weakly coupled to 
the rest of the system. 

It has to be stressed that one needs supersymmetry to naturally 
have flat directions and high temperature to lift them and 
stabilize their vevs at large nonzero values. The mechanism is 
very simple and natural because of the existence of a very large 
number of flat directions in the minimal supersymmetric standard 
model (MSSM) and other phenomenologically interesting models. 

{\bf Large external charge.} If one puts large enough charge 
in a system \cite{linde76,charge}, 
thermal excitations cannot `absorb' all of it and it must be `stored' 
in the vacuum, which thus becomes nontrivial. This is 
another easy way to give a nonzero vev to a scalar field. 
It is universal, being valid in both supersymmetric or 
ordinary models with gauge and/or local symmetries. 

What happens is that a scalar particle gets a negative mass 
term $-\mu^2|\phi|^2$ with $\mu$ the field's chemical potential. 
This term tends to give a nonzero vev to $\phi$, 
an opposite behavior with respect to the pure temperature 
contribution to the mass term, $+T^2|\phi|^2$. For a chemical 
potential (or, better, charge density) bigger than a critical one, 
the total mass term for $\phi$ becomes 
negative and $\phi$ acquires a nonzero vev, thus breaking some 
symmetry, since any field with a chemical 
potential must transform nontrivially under some group. 

However it is not a priori necessary that $\phi$ transforms nontrivially 
under exactly that symmetry, which originates the nonzero charge 
density. This is welcome, since one can achieve in this way a nonzero 
Higgs vev in the standard model at very high temperature 
with a large lepton number density 
in the universe although the Higgs boson does not carry a lepton 
charge \cite{linde76,liu94,brs97}! And this is exactly what 
could have happened in the early universe. The critical charge 
needed for the weak SU(2)$_L$ to be broken at any high enough temperature 
comes out to be of order $n_L\approx T^3$, i.e. of the order of 
the entropy density. Since the standard model has only one Higgs 
doublet, one cannot break also the electromagnetic U(1). This can be 
however easily achieved in the MSSM due to the presence of many 
scalars \cite{bs99}. 

There are still three issues we want to explain. First, 
such large lepton charges in the universe are allowed by 
the experimental data \cite{explepton}. 
Second, even if strictly speaking the lepton number itself is 
not conserved in the standard model, due to the breaking of 
weak SU(2)$_L$ sphalerons are not operative and the universe 
behaves similarly as it does at (almost) zero temperature today, 
i.e. effectively conserves lepton (and baryon) number \cite{liu94}. 
Third, the problem of producing such a large lepton number still 
remains. There has been some attempts \cite{casas97} in this 
direction using the Affleck-Dine mechanism, 
as well as possible explanations of the small baryon number 
\cite{mcdonald99}. 

\section{Cosmological applications}

The above ideas can be used in various mechanisms of baryogenesis 
\cite{baryo} and inflation \cite{infla} as well as to 
solve the following cosmological problems:

{\bf The monopole problem.} During a 
phase transition from a symmetry group G (take 
SU(5) for example) to a lower one H (the standard model 
SU(3)$_c\times$SU(2)$_L\times$U(1)$_Y$), monopoles are
created via the Kibble 
mechanism in many grand unified extensions 
of the standard model. Since monopoles created at the breaking 
scale of the grand unified theory would survive till today 
with at least ten orders of magnitude more energy density than 
baryons \cite{preskill79}, such a possibility is clearly 
unacceptable, and is referred to as the monopole problem. 
An elegant solution to this problem is to spontaneously 
break the initial group G or at least the U(1) factor in 
the final group H \cite{mohapatra79,langacker80,salomonson85,dms95}. 
This solution does not depend on the specific inflationary 
model used, and it does not pose any constraint on it, so 
that now the reheating temperature can be also large.

{\bf The domain wall problem.} This problem 
\cite{zeldovich74} is very similar in nature to the 
monopole one. 
The only difference is in the groups involved, so that it 
appears only when the vacuum manifold is disconnected. A 
typical example is a model with a discrete 
group $Z_2$. Again, too energetically domain walls get 
created during a phase transition between a phase with restored 
$Z_2$ (high $T$) and a phase with spontaneously broken 
$Z_2$ (low $T$). As before, if $Z_2$ is instead spontaneously 
broken at any temperature, there is no phase transition and 
thus no domain wall problem \cite{mohapatra79,ds95}. Such 
a solution is welcome whenever we have theories with 
spontaneously broken $P$, $CP$ or Peccei-Quinn symmetries. 

{\bf The false vacuum problem.} Take for example the case 
of supersymmetric SU(5). The effective potential 
has at low temperature three degenerate vacua, the SU(5) conserving, 
the one with the symmetry SU(4)$\times$U(1) and the standard model 
vacuum SU(3)$\times$SU(2)$\times$U(1), depending on the 
vev of the adjoint Higgs in the representation {\bf 24}. 
If one assumes SU(5) symmetry restoration at high temperature, 
the universe would remain in this same vacuum through all the 
history of the universe, since the barrier between different 
vacua are too high to tunnel through. If this were true, we could 
never live in our universe with the standard model symmetry group 
\cite{weinberg82}. Needless to say, a possible and elegant solution 
is given by symmetry nonrestoration: if 
SU(5) has been spontaneously broken at any high temperature to 
our standard model gauge group, no tunneling is necessary and 
the problem does not appear. 

\section*{Acknowledgments}
It is a pleasure to thank the organizers of COSMO99 as well 
as Mojca \v Cepi\v c, Gia Dvali, Alejandra Melfo, Toni Riotto 
and especially Goran Senjanovi\' c. This work was 
supported by the Ministry of Science and Technology of the 
Republic of Slovenia and by the Packard Foundation 99-1462 
fellowship.

\end{document}